# Synchronous Prediction of Arousal and Valence Using LSTM Network for Affective Video Content Analysis

Ligang Zhang and Jiulong Zhang
School of Computer and Engineering
Xi'an University of Technology
Xi'an, China 710048

*Abstract*—The affect embedded in video data conveys high-level semantic information about the content and has direct impact on the understanding and perception of reviewers, as well as their emotional responses. Affective Video Content Analysis (AVCA) attempts to generate a direct mapping between video content and the corresponding affective states such as arousal and valence dimensions. Most existing studies establish the mapping for each dimension separately using knowledge-based rules or traditional classifiers such as Support Vector Machine (SVM). The inherent correlations between affective dimensions have largely been unexploited, which are anticipated to include important information for accurate prediction of affective dimensions. To address this issue, this paper presents an approach to predict arousal and valance dimensions synchronously using the Long Short Term Memory (LSTM) network. The approach extracts a set of low-level audio and visual features from video data and projects them synchronously into pairs of arousal and valence values using the LSTM network which automatically incorporates the correlations between arousal and valance dimensions. We evaluate the performance of the proposed approach on a dataset comprising video clips segmented from real-world resources such as film, drama, and news, and demonstrate its superior performance over the traditional SVM based method. The results provide one of the earliest preliminary evidence to the benefit of considering correlations between affective dimensions towards accurate AVCA.

## I. INTRODUCTION

With the rapid proliferation of on-line video broadcasts via sharing platforms such as YouTube and Facebook, it becomes an increasingly challenging task to accurately reflect the semantics embedded in the video content for practical applications, such as semantic-based video retrieval and recommendation. As an intrinsic nature evolved from the origin of the human being, affect is one of the most important high-level semantics. The affect embedded in the video data can directly impact the perception and understanding of the content by reviewers. For instance, a comedy is likely to bring positive experience to reviewers, while a thriller film often imposes a fear or surprise feeling on reviewers.

Affective Video Content Analysis (AVCA) attempts to generate a direct mapping between video content and the corresponding affective states such as happiness or sadness. It can extract valuable insights from a piece of information about the emotional attitudes of the producer, and the entities involved (e.g. actors and activities) with respect to the topic portrayed in the electronic media. It also can help to summarize the key topics and content that are most likely to evoke similar affective responses from the reviewers from different resources such as news website, film, and social media. Being able to automatically identify the affective states in video content can be potentially used in many applications, such as video categorization, indexing, and recommendation.

Affect can be generally represented using discrete affective categories (e.g. surprise and sadness) or continuous affective dimensions (e.g. arousal and valence). The affective dimensions describe affective states using continuous axis values in a multi-dimension space. Compared with discrete categories, dimensional spaces are capable of representing a wider range of emotions, especially those spontaneous non-prototypical ones, as well as the correlation and intensity of emotions. Thus, they are more suitable for AVCA in real-world video data.

Generally, two types of methods are employed in existing studies on AVCA using affective dimensions: 1) constructing unsupervised models or knowledge-based rules to link the video content to affective dimensions by utilizing the knowledge learnt from associated fields such as psychology, art theory, and film [1], [2], [3], [4], and 2) employing supervised algorithms to learn the relationships between audio-visual data features and affective dimensions based on the training data [5], [6]. The first approach suffers from the drawback that the field knowledge is often subjective and may not fit for all types of real-world data. The second approach requires a well-designed machine learning strategy to learn inherently complicated correlations between video content and each dimension, as well as between multiple dimensions. Most existing studies project video features to each affective dimension separately, which has not taken into consideration the correlations between affective dimensions. However, such correlations are anticipated to also convey crucial information for accurate prediction of affect in video content.

To address the above issue, this paper presents an approach for AVCA that predicts affective dimensions synchronously based on a set of audio-visual video features using the Long Short Term Memory (LSTM) network. The intrinsic correlations between affective dimensions are inherently considered in the learning process of the LSTM network with multiple predicted outputs. We anticipate that considering the correlations between dimensions helps to improve the prediction of the relationships between audio-visual features and affective dimensions. The popular arousal-valance dimensional space is used here. Experiments on a video dataset collected from real-world resources confirm a superior performance of the proposed approach over a baseline Support Vector Machine (SVM) based approach which predicts each affective dimension separately.

The rest of the paper is organized as follows. Section II reviews prior related work. Section III introduces the proposed approach. Section IV presents experimental results. Conclusions are drawn in Section V.

## II. RELATED WORK

### A. Methods of Affect Representation

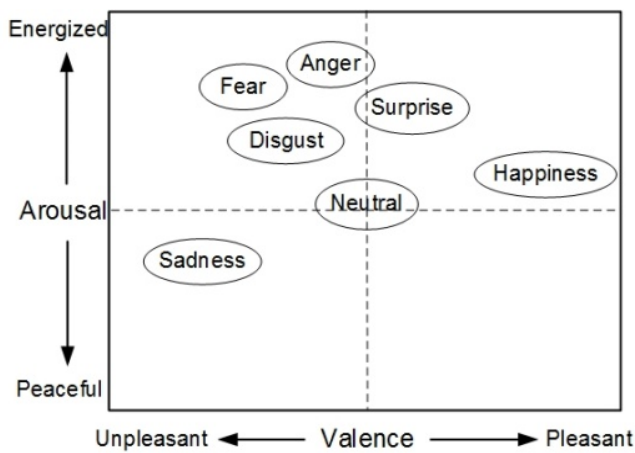

Figure 1. Distributions of six basic emotions plus neutral in the arousal and valence dimensional space.

There are primarily two types of ways of representing the affective states in video content: discrete affective categories and continuous affective dimensions. The discrete emotion category theory represents affect using one of pre-defined categories, such as six basic emotions - *anger (AN), disgust (DI), fear (FE), happiness (HA), sadness (SA), and surprise (SU)*, and non-basic emotions (e.g. interest, depression, and disagreement). One major issue of discrete categories is that not all emotions in the real-world video data can be represented by a limited number of categories with clearly pre-defined meanings, and it is still a challenge to precisely define a large number of various types of discrete categories. The dimensional theory describes emotions using continuous values in a multiple dimensional space, which was designed mainly based on psychological findings about human emotions. Examples of such affective dimensions are power, valence, activation and expectancy. Fig. 1 shows the widely adopted arousal-valence dimensional space [7] and the locations of six basic categorical emotions and neutral in this space. Compared with discrete emotion categories, affective dimensions have several advantages, including using values along a set of continuous axes to represent a wide range of emotions, especially those spontaneous non-prototypical ones in real-life data, providing unique insights into the intensity of emotions, and reflecting the similarity and contrast between categorical emotions.

### B. AVCA Approaches Using Affective Dimensions

There are generally two ways of reflecting the affective states embedded in video data – using either features extracted from the data content (e.g. color, texture, and motion), or emotional responses from viewers during watching data stimuli (e.g. facial expression [8], EEG signal [9], heart rate [10], and multiple physiological signals [11]). In this paper, we focus on approaches belonging to the first way that represents affective dimensions using features extracted from the data. For a recent survey on AVCA, readers are referred to [12], [13].

The vast majority of existing approaches to AVCA using affective dimensions are based on low-level features (e.g. keywords, lighting, motion, and color) [1], [2], [3], or middle-level features [5] generated based on low-level features. The correlation between extracted features and affective dimensions is built either from constructing models or rules by utilizing the knowledge learnt from psychology, art theory, color theory, aesthetics, and cinematography [2], [3], [4] etc., or from employing a supervised train-test process based on a training set using machine learning algorithms, such as neural networks [1], SVM [5], hidden Markov models, and fuzzy similarity [6]. These approaches primarily generate the mapping between audio-visual features and each affective dimension separately, and thus the intrinsic correlations between affective dimensions are largely ignored. As previous studies [7] and our previous work [14] showed that there often exist strong correlations between discrete affective categories and affective dimensions, as well as between affective dimensions, it is desirable to exploit the usefulness of the correlations between affective dimensions in improving the prediction accuracy of AVCA. The incorporation of such correlations is also anticipated to be an effective way to reduce the affective gap [2], which is caused by the lack of coincidence between the measurable features and the expected affective states of the viewer. We would like to highlight that this paper is one of the earliest studies on exploiting this issue.

## III. PROPOSED APPROACH

Fig.2 shows the proposed approach for predicting continuous affective dimensions associated with video content, which is composed of three main processing steps. For an input video clip, a set of representative audio and visual features, as well as their temporal statistical features are extracted which reflect the affective states embedded in the video content. These features are further reduced to a shorter dimension which keeps only the most effective features that are strongly associated with affective states using the Correlation-based Feature Selection (CFS) algorithm. The reduced features are further fed into the LSTM network, which projects the features to continuous affective dimensions synchronously in the arousal and valance space. The final results for each input video are predicted continuous values of arousal and valance dimensions, which can be used directly in applications such as affect-based video retrieval, indexing, and management.

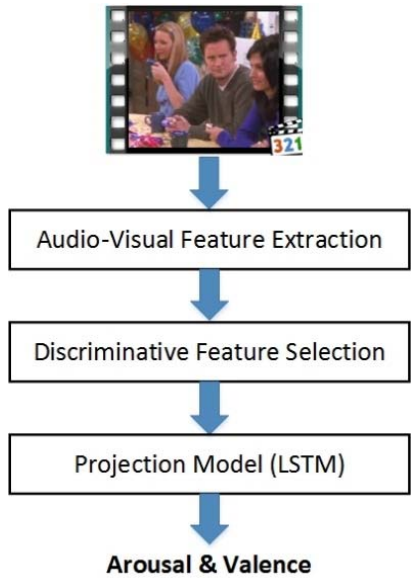

Figure 2. Framework of the proposed approach.

### A. Audio-visual Feature Extraction

The extraction of a set of effective features from video data is the first, but a critical step of the proposed approach. In the current literature, there is no consensus on what types of video features are most closely associated with the affective states in the data. However, it is generally believed that both audio and visual signals play important roles in conveying affective messages in the video content to viewers and making memorable affective impressions on viewers. In this paper, we extract a set of low-level audio and visual features based on the feature set that was both recommended by the Interspeech 2009 emotion recognition challenge [15], and well documented in existing studies that have close correlations with affective states. As shown in Table I, the audio features include pitch, energy, zero crossing rate, mel-frequency cepstral coefficients (MFCC 1-12), formant, and perceptual linear prediction, while the visual features are composed of motion intensity, lighting, color energy, shot switch rate, zero crossing rate, and rhythm regularity. In addition, clip-level temporal features (i.e. median, maximum, and minimum) of audio-visual features are also included to capture the temporal characteristics of affect. The openEAR open source implementation [16] is used for assisting the feature extraction.

TABLE I. AUDIO-VISUAL FEATURE SET USED FOR PREDICTING AFFECTIVE DIMENSIONS IN VIDEO CONTENT.

| | Feature |
|---|---|
| Audio | pitch, energy, zero crossing rate, mel-frequency cepstral coefficients, formant, perceptual linear prediction |
| Visual | motion intensity, lighting, color energy, shot switch rate, zero crossing rate, rhythm regularity |
| Temporal | median, maximum, minimum |

### B. Feature Subset Selection

Given a large set of audio-visual features, it is a common practice in automatic affective analysis systems to employ a feature selection algorithm to select only the most discriminative feature subset from the whole feature set. This is not only helpful in improving the performance of the proposed approach by utilizing only the most effective features, but also facilitates detailed result analysis and fast computational processing. In the current literature, there are many popular feature selection algorithms, such as Adaboost and linear discriminant analysis. However, these algorithms are mainly designed for categorical class classification problems and cannot be directly applied to the prediction task here, which actually involves handling of continuous affective dimensions. Thus, we choose to use the CFS algorithm [17], which is a simple, fast correlation based filter algorithm suitable for processing and predicting continuous regression problems. The CFS algorithm can be expressed mathematically as $Q_s = k\overline{r_{cf}}/\sqrt{k + k(k-1)\overline{r_{ff}}}$, where $Q_s$ is the merit of a feature subset $S$ containing $k$ features, $\overline{r_{cf}}$ and $\overline{r_{ff}}$ stands for the average feature-class correlation and the average feature-feature correlation, respectively.

During feature selection, the CFS algorithm begins with an empty feature set, then generates all possible single feature expansions at each search, and finally selects the subset with the highest evaluation. Among all audio-visual features, we keep only the top 20 features in the final feature set, which are further fed to the LSTM projection model for the prediction of affective dimensions.

### C. Affect Projection Using LSTM

A central task and main contribution of this paper is to investigate the benefit of synchronously prediction of arousal and valence dimensions based on a suitable machine learning algorithm. The aim of the projection model is essentially to map as accurately as possible the discriminative feature subset to continuous arousal and valance values. As shown in related work, in most current studies on AVCA, the mapping between video features and affective dimensions is generated separately for each dimension using rules or algorithms such as SVM. However, this mapping process accounts for only the relationship between features and affective dimensions, but suffers from the drawback of not considering inherent correlations between affective dimensions, which also include important information for dimension prediction because, in reality, dimensions in a specific affective space are always correlated to each other to a certain degree and a purely independent dimension very rarely exists.

In this paper, we adopt the LSTM network [18] as the projection model for mapping between video features and affective dimensions. There are two reasons for choosing this network: 1) it has a recurrent neural network architecture and can effectively utilize temporal features and correlations between video frames to achieve more accurate prediction of time series data. Affect often occurs across a certain length of time and is more effectively conveyed by temporal features such as sound, motion, action, etc. while static features in a single frame are often hard to elicit strong affective responses from reviewers. 2) It supports multiple inputs and outputs simultaneously and can predict the arousal and valance dimensions synchronously in the output layer, and thus it is suitable for the objective of this paper. The LSTM network has been successfully used for detecting emotions of viewers from EEG signals and facial expressions [9].

## IV. EXPERIMENTAL RESULTS

This section evaluates the performance of the proposed approach on an affective video dataset collected from real-world resources. We firstly describe the process of creating the video dataset and then present the prediction accuracy of the approach. The performance is also compared with a traditional SVM based approach.

### A. Data Collection

To evaluate the performance of the proposed approach in predicting affective dimensions in real-world data, a naturalistic affective video dataset was created from real-world resources, including films, talk shows, news, dramas, sports, etc. The criterion used for video selection is that the video content is supposed to elicit strong affective responses from viewers. Various types of films including comedy, action, disaster, horror, etc. are included that are generally believed to contain different types of strong affective content. From a whole video sequence, we manually segmented shorter clips that have only one dominant affective state, such as happiness or sadness, in each clip. Then three subjects (postgraduate candidates majoring in computer science) were recruited to manually annotate arousal and valance dimensions of each video clip to one of five categorical levels, i.e. [-2, -1, 0, 1, 2], using the widely adopted Self-Assessment Manikins method. Take the valance dimension as an instance, -2 indicates that there is very unpleasant affect in the video content, while 2 stands for the strongest pleasant affect. To remove possible disagreements in the annotation results between subjects, the average values of three subjects are used as the results. An expert in affect analysis further validated and assigned the results to a closet categorical level, and any confusion results were removed accordingly. The resulting dataset includes 200 video clips with varied frame resolutions and durations ranging from 5 to 30 seconds. Fig. 3 shows screenshot examples of the video clips and their annotations.

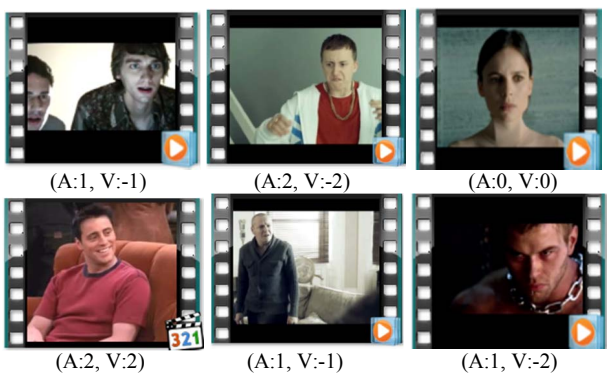

Figure 3. Screenshot samples of video clips and their corresponding annotated arousal (A) and valance (V) values.

### B. Experimental Settings

*Performance metric*: four metrics are used to measure the prediction performance of the proposed approach, including:

1) $R^2$ statistic, which measures the proportion of the variation of the observations around the mean. A value of 1 means a perfect fit between the prediction results and ground truths, while a negative value means inaccurate prediction.

2) Pearson Correlation Coefficient (CC), which measures the strength of a linear relationship between two variables. 1 means perfect correlation, while 0 implies uncorrelated.

3) Mean Linear Error (MLE), which is the average of the absolute error between the predicted results and ground truths.

4) Bhattacharyya Distance (BD), which measures the similarity of two probability distributions. 0 means that two distributions are similar, while a larger value indicates a bigger difference.

*Evaluation strategy*: four random cross-validations are used to obtain the prediction results. All video clips are randomly divided into four folds, and in each validation, three folds are used for training and the rest one fold for testing. This process repeats four times to generate the final results.

### C. Prediction Results

Table II shows the prediction results of the proposed approach. From the table, we can see that using a fusion of audio and visual features leads to better overall performance than using audio or visual alone for all four measurements, including $R^2$, CC, MLE, and BD. This confirms the general knowledge that both audio and visual signals contain useful information in conveying affective content in real-world video clips and they can complement each other to improve the accuracy of predicting arousal and valence dimensions. There are not big differences in the performance results between audio and visual features, indicating that they are approximately equally important for affective dimension prediction on the evaluation dataset. Prediction of valence seems to be easier than that of arousal using the proposed approach, which has higher $R^2$ and CC as well as a lower BD for predicting valence than predicting arousal. This is probably because the discrimination of unpleasantness vs. pleasantness (i.e. valance) is easier than the identification of the intensity of affective states (i.e. arousal) using machine learning algorithms, which also agrees with the human's recognition of affective emotions. However, it is interesting to observe that arousal has a lower MLE than valance, and this verifies the fact that different measurement metrics are able to reflect different aspects of the prediction results.

TABLE II. PREDICTION RESULTS OF THE PROPOSED APPROACH.

| Feature | Arousal | | | | Valence | | | |
|---|---|---|---|---|---|---|---|---|
| | R2 | CC | MLE | BD | R2 | CC | MLE | BD |
| Audio &Visual | 0.350 | 0.575 | 0.245 | 0.043 | 0.595 | 0.711 | 0.385 | 0.025 |
| Audio | 0.210 | 0.537 | 0.321 | 0.052 | 0.514 | 0.684 | 0.416 | 0.031 |
| Visual | 0.252 | 0.542 | 0.288 | 0.050 | 0.463 | 0.645 | 0.420 | 0.025 |

Fig. 4 visually illustrates the position distributions of a sample set of video clips in the arousal and valance space, based on the prediction results of the proposed approach. The coordinates represented by the predicted arousal and valance values are shown together with the representative frame in the corresponding video clip. Based on these samples, we can clearly see that the proposed approach achieves promising results in identifying unpleasantness vs. pleasantness, and low vs. high intensity of affective states in video clips. The unpleasant video clips can be easily distinguished from pleasant ones, while the intensity of emotions can also be identified clearly based on the prediction results.

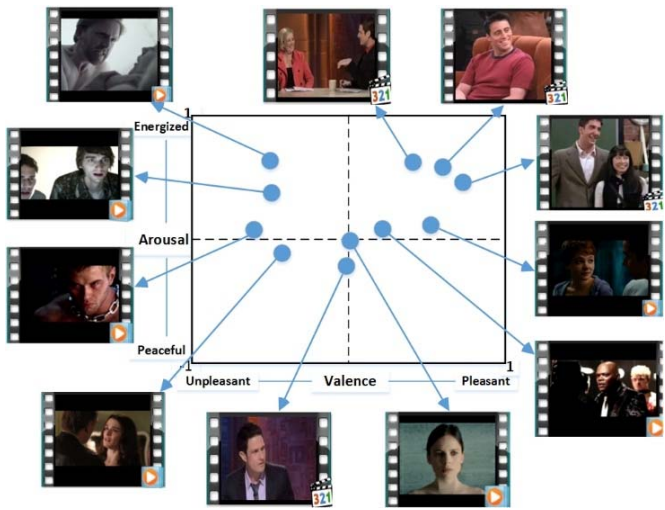

Figure 4. Distribution of sample video clips in the arousal and valance space based on the prediction results of the proposed approach.

### D. Performance Comparisons with SVM based Approach

An important question is whether considering intrinsic correlations between arousal and valance dimensions indeed leads to a better performance compared to without considering the correlations? To answer this question, we choose an SVM based approach as the baseline for performance comparisons. The SVM based approach takes as input the same audio and visual feature set as in the proposed approach, but performs the prediction of arousal and valance dimensions separately. Thus the intrinsic correlations between arousal and valance are not considered in the SVM based approach.

TABLE III. PERFORMANCE COMPARISIONS OF THE PROPOSED LSTM BASED APPROACH WITH SVM BASED APPROACH.

|  | Arousal | | | | Valence | | | |
| --- | --- | --- | --- | --- | --- | --- | --- | --- |
|  | R2 | CC | MLE | BD | R2 | CC | MLE | BD |
| LSTM | 0.350 | 0.575 | 0.245 | 0.043 | 0.595 | 0.711 | 0.385 | 0.025 |
| SVM | 0.286 | 0.520 | 0.362 | 0.055 | 0.513 | 0.615 | 0.400 | 0.124 |

Table III shows performance comparisons of the proposed LSTM based approach with the SVM based approach. It is obvious to see that the proposed approach has a better overall performance than the SVM based approach for all four measurements. To be specific, the proposed approach has higher $R^2$ and CC, and lower MLE and DB than the SVM based approach. The results confirm our initial assumption that the intrinsic correlations between arousal and valance are actually useful in the prediction of affective content in video clips. The results also prove the motivation of designing the proposed approach and provide experimental evidence to the benefit of considering correlations between affective dimensions in designing affective prediction algorithms.

## V. CONCLUSIONS

This paper presents a LSTM based approach to synchronously predict affective dimensions of real-world video clips in the popular arousal and valance space. A major contribution lies in exploiting the usefulness of considering intrinsic correlations between arousal and valance dimensions to improve the performance of predicting dimensional affect embedded in video content. The performance of the proposed approach is measured by four metrics on a realistic affective video dataset with manually annotated ground truths of arousal and valance dimensions. The experimental results show that both audio and visual features are approximately equally important for predicting affective dimensions and fusion of them leads to a better performance than using each alone. Performance comparisons to a traditional SVM based approach, which predicts arousal and valance dimensions separately and does not consider correlations between them, confirm a superior performance by considering correlations between affective dimensions. Our future work will conduct comparisons with more prediction algorithms, such as feed-forward neural networks, and tests on more real-world video data, such as the 'in the wild' dataset [19].

ACKNOWLEDGMENT

This work is supported by the National Natural Science Foundation of China (No. 61402362), Natural Science Foundation of Shaanxi (No. 2015JQ6218), Shaanxi Education Bureau Science (No. 16Jk1553), and Beilin Bureau Science (No. GX1616). The authors thank three subjects for annotating ground truths of the naturalistic affective video dataset.